\documentclass{elsart}
\usepackage{natbib}
\usepackage{graphicx}
\begin{document}
\begin{frontmatter}
\title{The Belle Silicon Vertex Detector}
\author{Hiroaki Aihara}
\address{Department of Physics, University of Tokyo, Tokyo 113-0033, Japan}
\begin{center}
{\small \it (To appear in the Proceedings of the 4th International Symposium on Development and Application of Semiconductor Tracking Detectors, March 22 - 25, 2000 Hiroshima, Japan )}
\end{center}
\begin{abstract}
The silicon vertex detector of the Belle experiment
has been  built  to provide the vertex information of $B$ meson decays
at the KEKB accelerator.
It consists of three cylindrical layers of double sided, double metal
silicon sensors read out by VA1 chips. The analog signals are transmitted 
to a system of fast analog-to-digital converters and are reduced on
a system based on Motorola Digital Signal Processors.
Performance during the initial data taking period and a brief 
description of the upgrade plan are also presented.

\end{abstract}
\end{frontmatter}

\section{Introduction}

The Silicon Vertex Detector (SVD) of the Belle experiment~[1] at the KEKB 
accelerator~[2] has been built to provide vertex information from precision
measurements of charged tracks in the vicinity of  the interaction point.
Because the KEKB is an asymmetric $e^+e^-$ collider with 8~GeV $e^-$  
and 3.5~GeV $e^+$ beams,
a pair of $B$ and $\overline{B}$ mesons, produced nearly 
at rest at the
center of mass system (CMS),  is  boosted along the $e^-$ beam direction
with a Lorentz boost factor of $\beta\gamma=0.425$.
Therefore, in designing  the SVD a special emphasis is placed on 
the precise measurement of the $z$ 
(boost direction) coordinate.
The typical separation between $B$ and $\overline{B}$ decays at the KEKB is
$\sim 200~\mu$m and we find, based on the simulation,
the $z$ resolution of $\sim 100~\mu$m
in $B$ meson decay vertex measurement is sufficient to detect a possible
asymmetry in the proper time interval distribution 
for decays of $B^0-\overline{B^0}$ pairs. 
This asymmetry  signals $CP$-violation in the $B$ meson system 
when one of the neutral $B$ mesons decays to a $CP$ eigenstate.
\par
\begin{figure}[thb]
\begin{center}
\includegraphics[scale=0.7,clip]{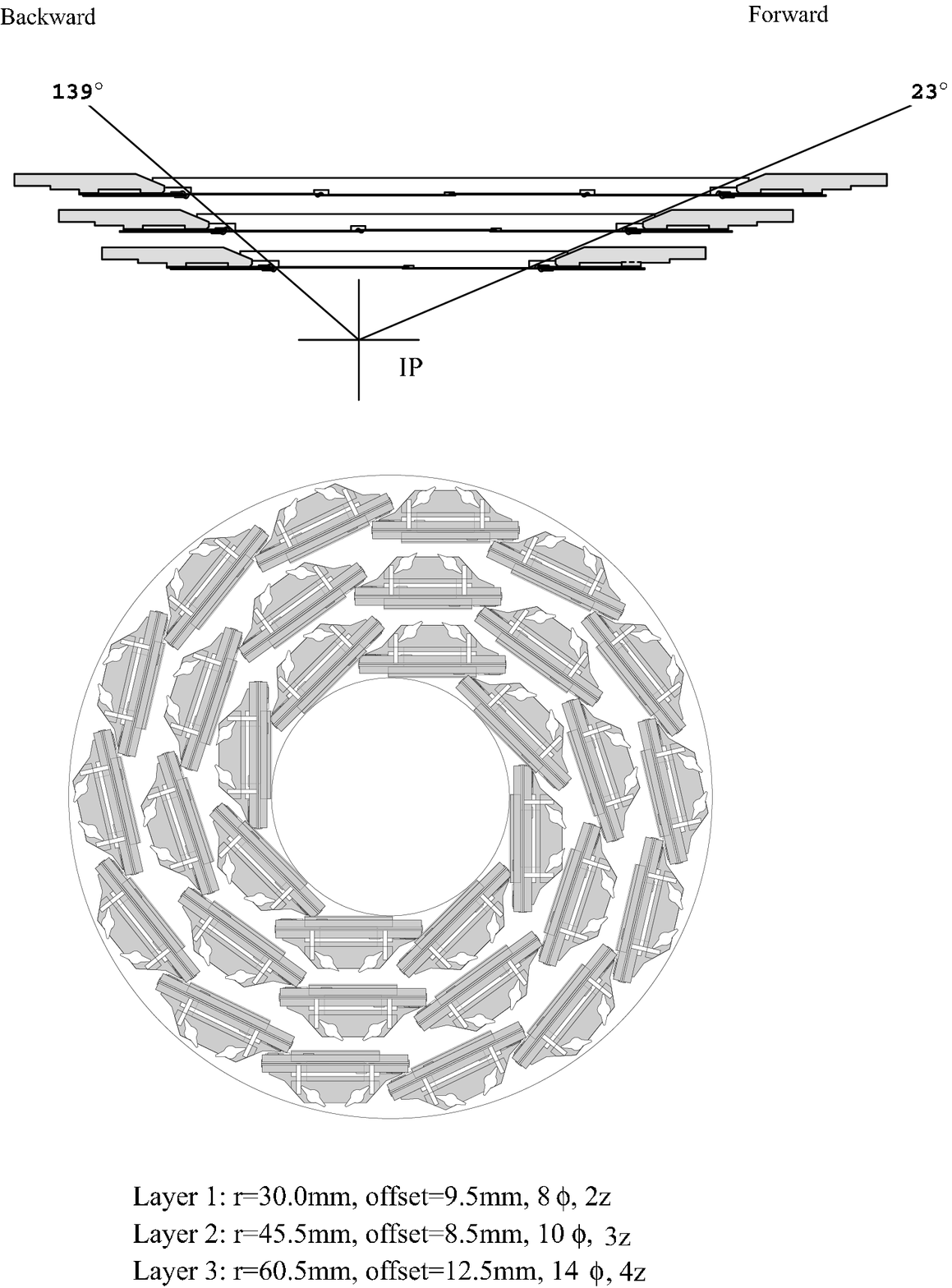}
\caption{Geometrical layout of the Belle SVD
\label{fig1}}
\end{center}
\end{figure}

The SVD consists of three concentric cylindrical layers of silicon sensors
with two orthogonal-coordinate readouts.
The vertex resolution for low momentum tracks is dominated by 
multiple scattering in a material before the SVD plus the first SVD layer.
Figure~\ref{fig1} displays the geometrical layout of the SVD.
The radii of the three layers are $30$~mm, $45.5$~mm and $60.5$~mm. 
The SVD covers the polar angle range of $23^\circ<\theta<139^\circ$
corresponding to $86\%$ of the solid angle in the CMS.
There installed is
a $20.5$~mm-radius beryllium beampipe  with two $0.5$~mm-thick walls
allowing helium gas flow to cool the
beampipe.
A 20~$\mu$m gold foil is glued on the inner concentric 
cover consisting of carbon-carbon material to protect the SVD from
soft Xrays produced in the beampine.
The material before the first SVD layer amounts to $0.95\%$ of 1 radiation
length.
\par
\begin{figure}[thb]
\begin{center}
\includegraphics[scale=0.6,clip]{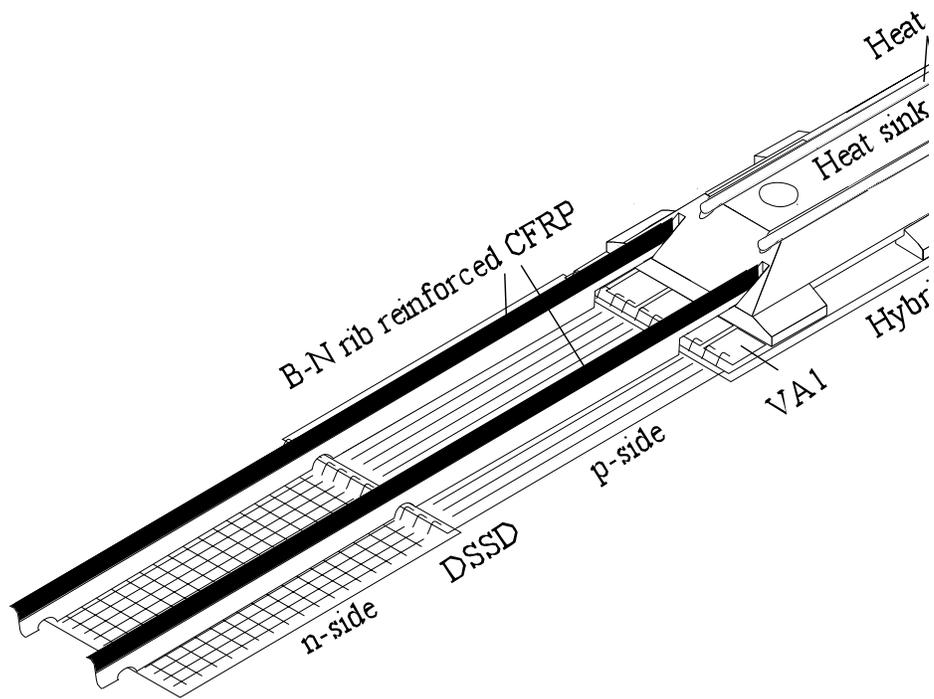}
\caption{Construction of the SVD ladder. In order to form a two-DSSD unit,
n-strips of one DSSD are wire-bonded to p-strips of the other.
\label{fig2}}
\end{center}
\end{figure}

The first layer is formed from 8 ``ladders'', each of which contains
two double-sided silicon strip detectors (DSSDs) which are read out separately.
The second layer is formed from 10 ladders, each containing  3 DSSDs.
Two of 3 DSSDs are connected to make a single two-sensor unit, in which
 p-strips on one DSSD are wire-bonded 
to n-strips on the other DSSD to minimize a noise in a  channel.
The third layer is formed from 14 ladders, each containing 2 two-sensor units.
A full ladder consists of DSSDs, aluminum nitride hybrids at each end, 
glued with
an aluminum nitride heat sink containing two heat pipes and 
CFRP reinforced boron nitride ribs glued across DSSDs and the heat sink,
as shown in Fig.~\ref{fig2}.
The DSSD, HPK S6936, was originally fabricated for the DELPHI vertex detector
by Hamamatsu Photonics.
It has an active area of $54\times 32~{\rm mm}^2$ with a thickness of $300~\mu{\rm m}$.
The strip pitch for a p(n)-side is $25(42)\mu$m and 
the readout pitch is $50(82)\mu$m.
This DSSD contains integrated coupling capacitors and, therefore,  
the strips are  AC-coupled to 
preamplifiers.
It utilizes a double metal structure to read out n-strips.
The total number of readout channels of the SVD is 81920.
A bias voltage of 75V is supplied on n-side.
\par
Each hybrid is made by gluing two identical aluminum nitride boards,
each containing five 128-channel VA1 chips~[3] providing 640
readout channels per board.
The VA1 chip fabricated by 1.2~$\mu$m feature size process
contains preamplifiers, shapers and  track-and-hold circuits
that capture the analog pulse-height information of each channel.
The peaking time of the shaper is set to $1.0~\mu$s
to match the trigger latency.
When the system switched from track- to hold-mode the stored analog 
information for each channel is sequentially routed via an on-chip analog
multiplexor to a system of fast analog-to-digital converters (FADCs)~[4].
Event buffering and zero data suppression are performed 
using Motorola Digital Signal Processors (DSPs) in the FADC module. 

\section{Basic Performance}

\begin{figure}[tbh]
\begin{center}
\rotatebox{-90}{\includegraphics[scale=0.8,clip]{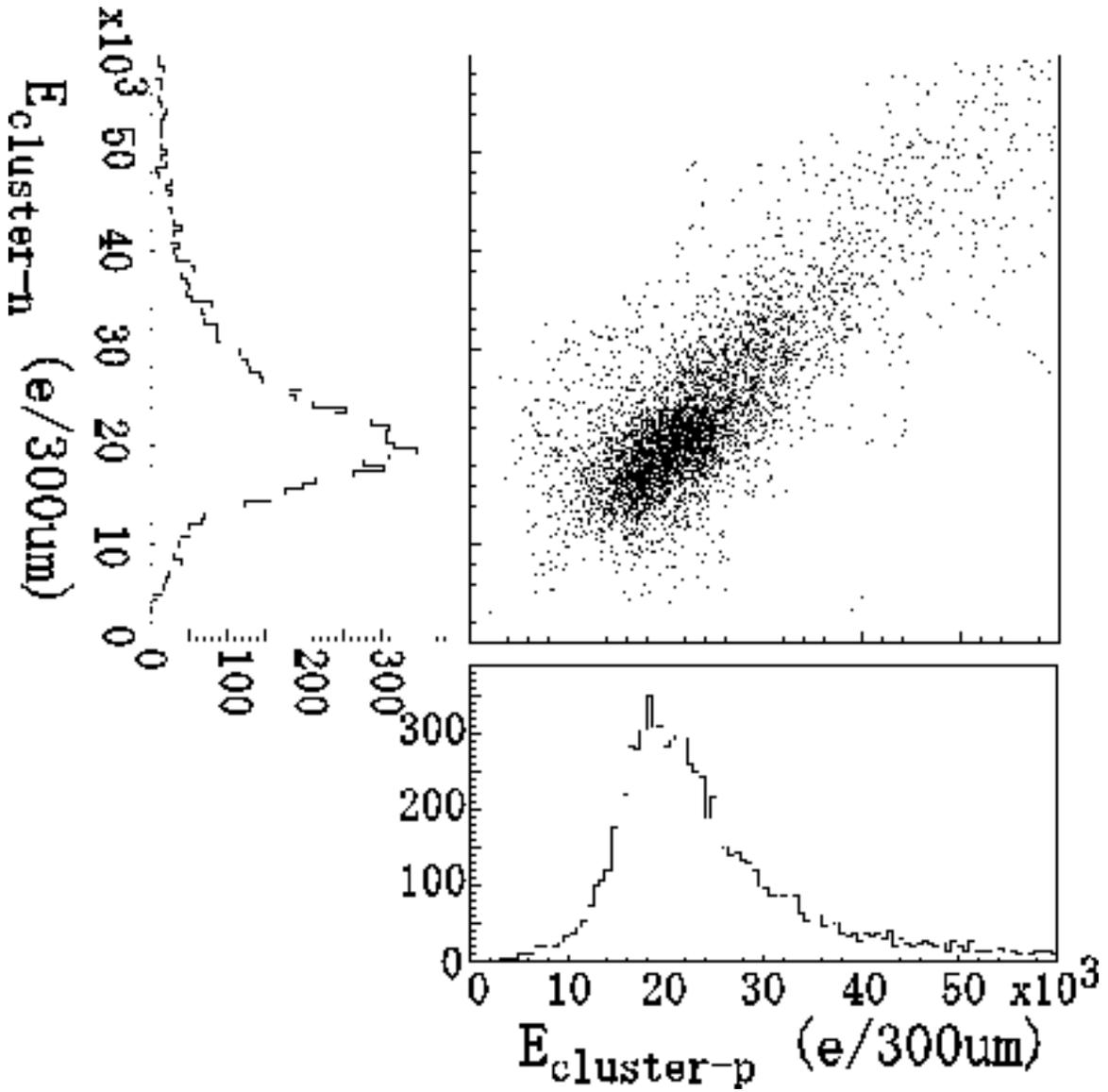}}
\caption{ $dE/dX$ distributions per $300~\mu{\rm m}$-thick silicon of  minimum ionizing particles 
given in units of electrons.
\label{fig3}}
\end{center}
\end{figure}

The $dE/dX$ distributions of minimum ionizing particles 
for p-side and n-side,
together with the correlation between signals from two
sides  are shown 
in Fig.~\ref{fig3}.
The $dE/dX$ values are calculated by summing all signals (above threshold) 
which belong to the same track and by normalizing 
to the same track length (i.e.,
300~$\mu$m).
The signal peaks at $\sim 18,800e$ and $\sim 19,200e$ for p-side and n-side, 
respectively, while
the electronics noise is found to be $\sim 550e$ for p-side channels,
$\sim 1000e$ for n-side channel and $\sim 1100e$ for p-n combined channels
in which a p-strip and an n-strip are connected.
Therefore, Signal-to-Noise (S/N) ratio is well above 10.
The strip yield for each layer is found to be $98.8\%$, $96.3\%$ and $93.5\%$
for 1st, 2nd and 3rd layers, respectively.
The loss of strips includes failure of hybrids and 
a few failed coupling capacitors.
Since the installation in August 1999, the SVD has been functioning stably.
The occupancy of the SVD is $\sim 1\%$ at the luminosity of 
$1.2\times 10^{33}{\rm cm}^{-2}{\rm s}^{-1}$.
For hadronic events we find $97\%$ of tracks reconstructed by the Central Drift Chamber (CDC)  
have corresponding hits
in the SVD.  
The internal alignment of the DSSDs and the 
SVD-CDC relative alignment have been  performed
using cosmic rays taken with and without beams in the KEKB.
In order to illustrate the so-far achieved alignment we summarize in Table~1 
resolutions of track helix parameters\footnote{
A charged particle track is expressed as: $x=x_0+d\rho\cos\phi_0+\frac{\alpha}{\kappa}
[\cos\phi_0-\cos(\phi_0+\phi)]$, $y=y_0+d\rho\sin\phi_0+\frac{\alpha}{\kappa}[\sin\phi_0
-\sin(\phi_0+\phi)]$, $z=z_0+dz-\frac{\alpha}{\kappa}(\tan\lambda)\phi$, 
$\alpha\equiv 1/cB({\rm Tesla})$.}
compared with the values for the perfect
alignment,  predicted by Monte Carlo simulation.
We note that 
the $dz,\tan\lambda$ resolutions are mostly determined 
by the SVD internal alignment 
performance, while
the $d\rho,\phi_0,\kappa$ resolutions depend on 
the  CDC track resolutions as well.

\begin{figure}[thb]
\begin{center}
\includegraphics[scale=0.7,clip]{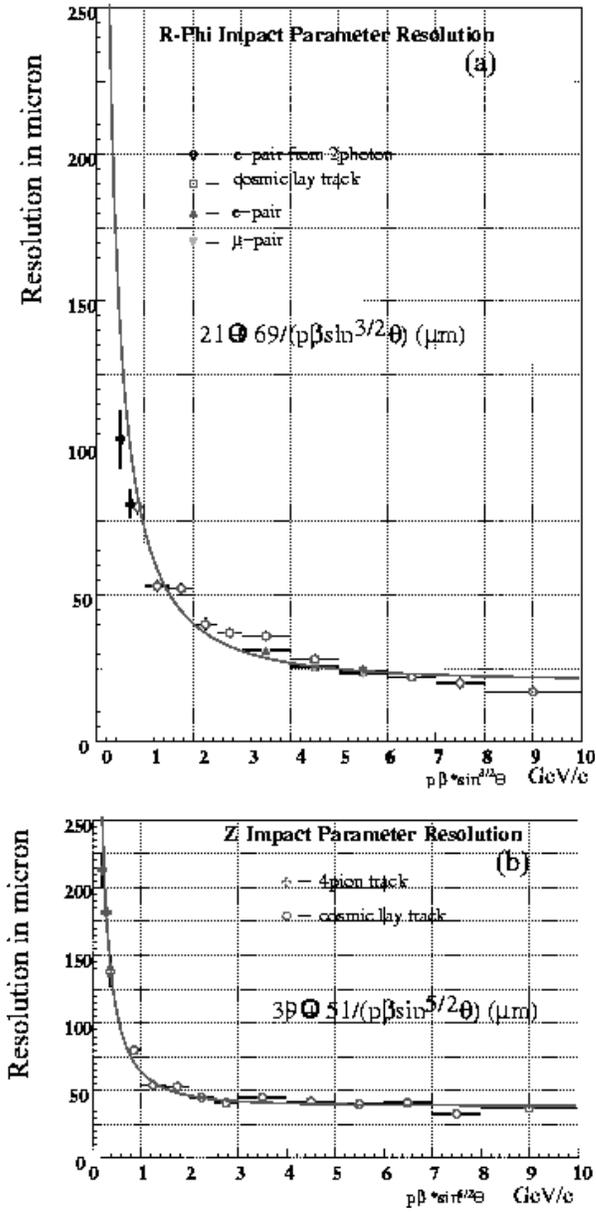}
\caption{Impact parameter resolutions for (a) $r\phi$  and (b) $z$ directions
\label{fig4}}
\end{center}
\end{figure}

Figure~\ref{fig4} shows impact parameter resolutions, 
$\sigma_{r\phi},\sigma_z$,
for $r\phi$ and $z$ directions
measured using 
cosmic rays, Bhabha events and $\gamma\gamma\to e^+e^-,\pi^+\pi^-\pi^+\pi^-$ processes.
We obtain
$$
\sigma_{r\phi} =  \sqrt{(21)^2+\left(\frac{69}{p\beta\sin^{3/2}\theta}\right)^2}\ \mu{\rm m};\ \ \ \ 
\sigma_z       =  \sqrt{(39)^2+\left(\frac{51}{p\beta\sin^{5/2}\theta}\right)^2}\ \mu{\rm m},
$$
where $p$ and $\beta$ are momentum in ${\rm GeV}/c$ and 
velocity divided by $c$ of the particle, 
and $\theta$ is the polar angle from the beam axis.
The resolution in a measurement of the distance between 
two $z$ vertices of leptons originating from a $J/\psi$ particle can also 
infer the resolution of the $z$ coordinate measurement.
Here, each $z$ vertex is calculated by finding an intersection 
between a lepton track and the beam axis.
We find $\sigma_{\Delta z(\ell^+\ell^-)}=138\mu {\rm m}$, 
while Monte Carlo predicts $112\mu{\rm m}$.

\section{Physics Examples}
\begin{figure}[thb]
\begin{center}
\includegraphics[scale=0.7,clip]{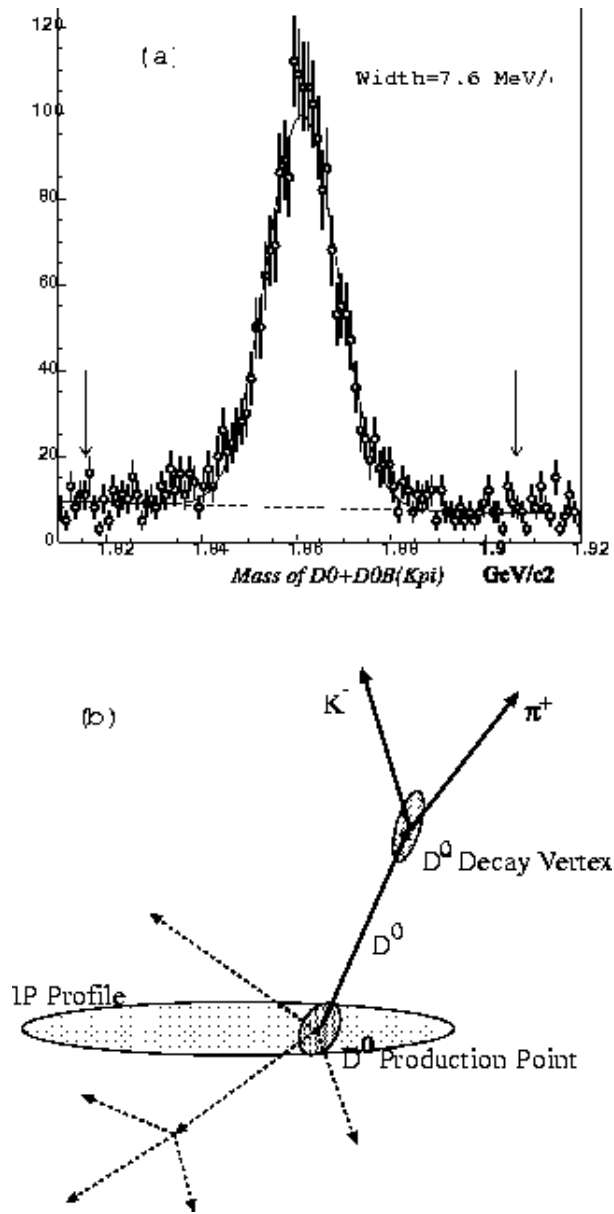}
\caption{(a)Invariant mass distributions of 
$D^0\rightarrow K^-\pi^+$
decays. (b)Illustration to show  how the $D^0$ meson flight length is
reconstructed in the plane perpendicular to the beam axis.
\label{fig:d0mass_fig}}
\end{center}
\end{figure}

In order to demonstrate the overall performance of the SVD, 
we present preliminary 
results of $D$ and $B$ meson lifetime measurements based on the first
$0.25\ {\rm fb}^{-1}$ data collected during the  initial data taking period~[5].
We reconstruct $D^0$ mesons in the decay mode $D^0\rightarrow K^-\pi^+$.
Charged tracks forming a $D^0$ candidate are required to be 
well-reconstructed
in the tracking system
and to originate from a common vertex.
Kaons are positively identified in the particle ID system~[6].
In order to avoid bias due to $B$ meson lifetime,
$D^0$ mesons originating from $B$ mesons
are removed by requiring
$D^0$ mesons  have a minimum momentum of $2.5~{\rm GeV}/c$
in the $\Upsilon(4S)$ rest frame.
Figure~\ref{fig:d0mass_fig}(a) shows the 
invariant mass distribution of $D^0\rightarrow 
K^-\pi^+$
decays.
By fitting the distribution with a 
Gaussian for the signal plus a linear function
for the background, we find
the width of the signal distribution to be  $7.6\pm 0.2~{\rm MeV}/c$
and the number of signal events to be $1730\pm 50$.
The background fraction is calculated from the same fit and is about $10 \%$
under a signal peak (the region within $\pm 3 \sigma $).

\begin{figure}[thb]
\begin{center}
\rotatebox{-90}{\includegraphics[scale=0.7,clip]{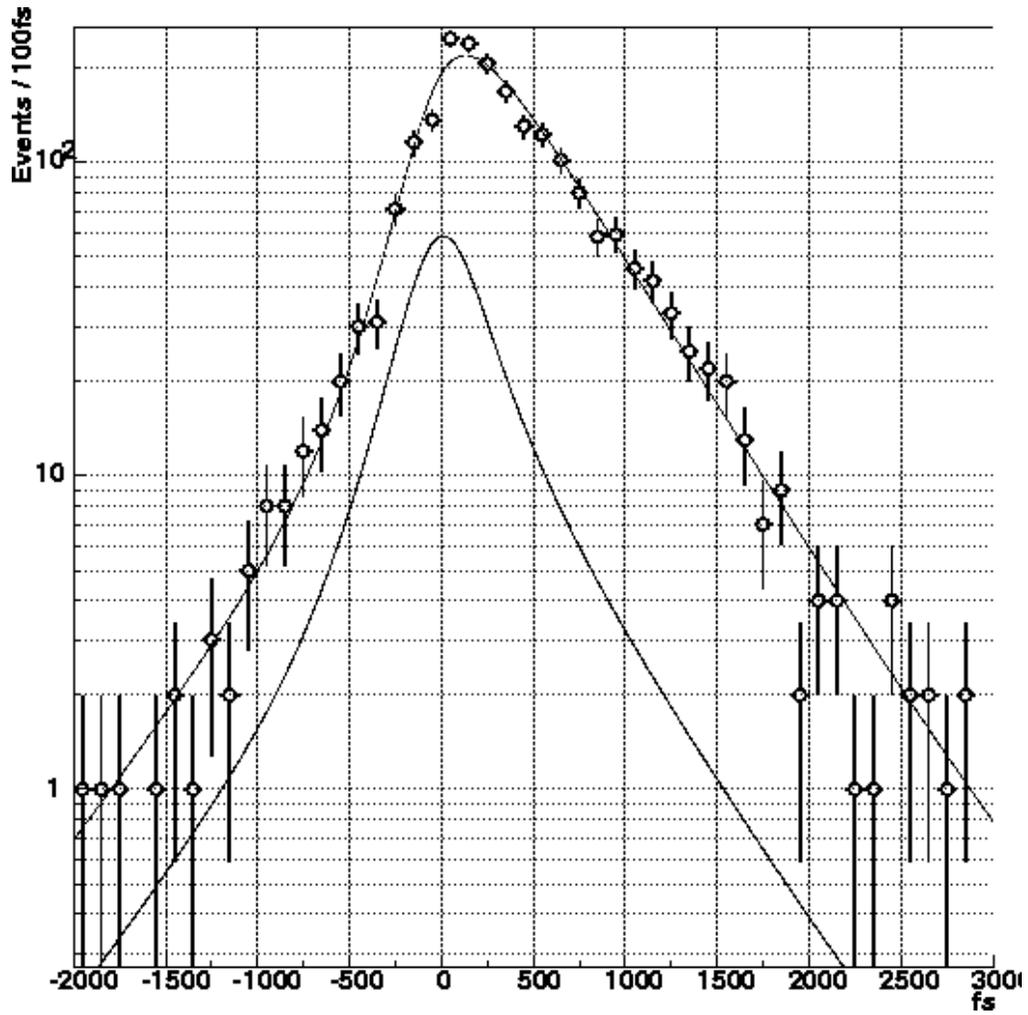}}
\caption{The proper-time distribution of the $D^0$ mesons
superimposed with the fit result.
The solid line with data points is the overall fit result and
the solid line without data points indicates the contribution from
the background.
\label{fig:d0fit}}
\end{center}
\end{figure}

As illustrated 
in Fig.~\ref{fig:d0mass_fig}(b) the lifetime of the charmed meson
is computed from the flight distance between the reconstructed $D^0$
vertex and the interaction point(IP) determined for each KEKB beam fill.
The profile of the IP is known to have, from the machine optics
and the beam size monitoring information,
$\sigma_z\sim 4~{\rm mm}$ along the beam direction  $(z)$, and 
$\sigma_x\sim 100~\mu{\rm m}$, and $\sigma_y\sim 5-10~\mu{\rm m}$ 
in the direction perpendicular to the beam direction.
The $D^0$ flight length is measured in the $xy$ plane to take advantage 
of the small IP size in the  plane.
The projected decay length $\ell_{decay}^{xy}$ of the $D^0$ meson is
calculated from the $D^0$ momentum ${\bf p}^{xy}_{D^0}$, 
the $D^0$ decay vertex ${\bf r}^{xy}_{D^0}$, and 
the $D^0$ production point ${\bf r}^{xy}_{IP}$ 
reconstructed by extrapolating the $D^0$ momentum
back to the IP profile:
$\ell^{xy}_{decay}=({\bf r}^{xy}_{D^0}-{\bf r}^{xy}_{IP})
\cdot{\bf p}^{xy}_{D^0}/|{\bf p}^{xy}_{D^0}|$.
The proper time $t$ is, then, calculated using the $D^0$ mass, $m$, as
$t = (m/c)(\ell^{xy}_{dec}/|{\bf p}^{xy}_{D^0}|)$.
The $D^0$ meson lifetime is extracted from the proper-time distribution 
with an unbinned maximum likelihood method.
The detector resolution is mostly described by a single Gaussian 
with a small correction for events which might have undergone a hard multiple 
Coulomb scattering and/or 
have  wrong CDC-SVD hit associations in the track reconstruction program.
Figure~\ref{fig:d0fit} shows the fit to the proper time distribution.
The preliminary result for the lifetime of $D^0$ meson is
$405.3\pm 13.3\ {\rm fs} ,$
where the error includes only the statistical error.  
This is in good agreement with the PDG value of $415\pm 4$~fs.

\begin{figure}[tbh]
\begin{center}
\includegraphics[scale=0.7,clip]{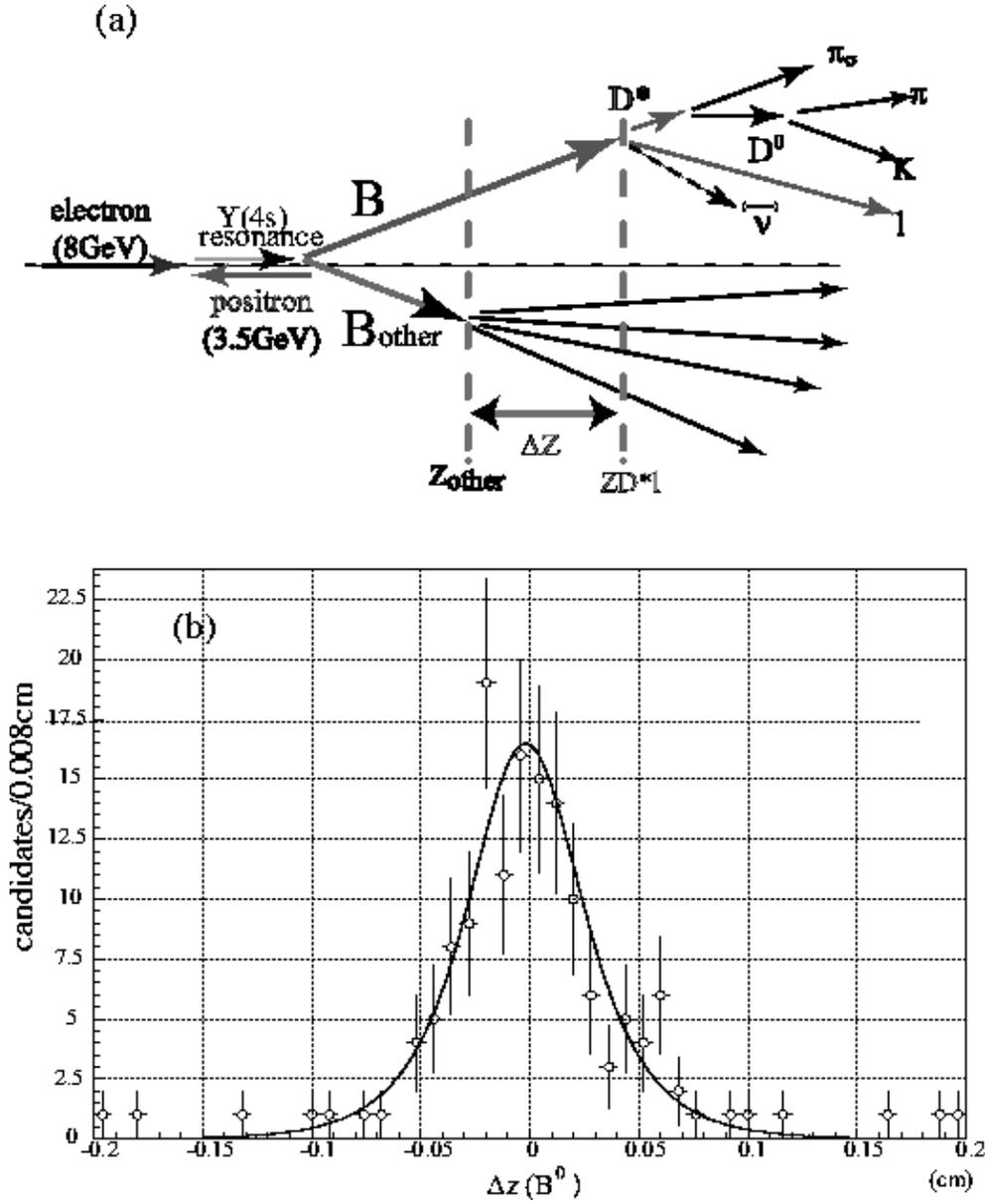}
\caption{(a) Illustration of reconstruction of 
$B$ and $\bar{B}$ decay vertices.
(b) Separation of $B$ and $\bar{B}$ decay vertices in the $z$ direction.
\label{brecon}}
\end{center}
\end{figure}

The $B$ meson lifetime is extracted from the measurement of the separation
of $B$ and $\bar{B}$ decays, $\Delta z$, in the $z$ direction.
We first reconstruct $B^0\to D^{*-}\ell^+\nu$, where $D^{*-}$ decays to 
$\overline{D^0}\pi^-$ followed by either one of $\overline{D^0}\to
K^+\pi^-$, $K^+\pi^-\pi^0$, $K^+\pi^-\pi^-\pi^+$ decays.
The decay vertex of $B^0$ is determined  using the reconstructed 
$\overline{D^0}$ momentum and the lepton $(\ell^+)$ track.
We, then, determine the $\overline{B^0}$ decay vertex requiring
all the remaining tracks in the same event, except for  
tracks originating from $K_S$ decays and tracks with large impact 
parameters, form a common vertex.
Figure~\ref{brecon}(a) illustrates the vertex reconstruction of $B-\bar{B}$
events.
The distribution of $\Delta z$ with the fit result is shown 
in Fig.~\ref{brecon}(b).
The fitted function consists of the expected $\Delta z$ distribution smeared
with the detector resolution.
We obtain the result of $\tau_{B^0}=1.82\pm 0.28$~ps, while the resolution
in $\Delta z$ is found to be $\sigma=146\pm 47~\mu{\rm m}$.
The result is consistent with the PDG value of $\tau_{B^0}=1.56\pm 0.04$~ps.
\section{Upgrade Plans}
The current SVD readout chips have limited radiation
tolerance.
The  $1.2~\mu{\rm m}$-technology VA1 chip is not based on  a radiation-hard technology and
stops functioning after a $\gamma$-ray dose of $\sim 200$~krad.
With the current radiation  level  observed at the KEKB, this corresponds to
an integrated luminosity of $\sim 8~{\rm fb}^{-1}$. 
We have been  developing  a frontend IC which stands radiation dose of
more than 1Mrad by taking advantage of the thinner 
gate-oxide layers used in sub-micron CMOS processes.
Because the threshold shift in a MOS transition  scales  as the square
of the thickness of the gate oxide layer, 
which in turn is 
proportional to the FET channel width  or  equivalently to the feature
size of the process being used,
one expects the radiation tolerance of the IC scales as 
the inverse square of the feature size.
One expects that  a $0.5~\mu$m process IC is four times more 
radiation-tolerant than a $1.0~\mu$m process IC.
We have fabricated two versions of VA1 chips, one with AMS $0.8~\mu$m technology
and the other with AMS $0.35~\mu$m~[7].
Figure~\ref{va08}(a) shows gain versus radiation dose for AMS $0.8~\mu$m
technology VA1. The gain decrease rate is found to be $-0.06\%$/krad.
Figure~\ref{va08}(b) shows the equivalent noise count versus  radiation dose.
Based on these measurements we conclude the $0.8~\mu$m VA1 chips
are radiation tolerant up to $\sim 1$Mrad.
The new SVD built with these VA1 chips is scheduled to be installed in summer
2000.
 The AMS $0.35~\mu$m VA1 exhibits the excellent radiation hardness as 
shown in Fig.~\ref{va035}. It is tolerant up to 20~Mrad.
The SVD based on this $0.35~\mu$m VA1 chips is also being planned.

\section{Conclusion}
We have successfully  built the SVD for the Belle experiment at the KEKB. 
Preliminary measurements of lifetimes of $D^0$ and $B^0$ mesons 
show it is performing as designed.
We have developed radiation-hard VA1 chips, based on a sub-micron 
process technology, which will be applied
to the immediate upgrade of the SVD.

\section{Acknowledgments}
The author is grateful to the conference organizers for their hospitality,
and indebted to colleagues of the Belle SVD group for their help in preparing
the talk. 
This work is supported by a Grant-in-Aid for Scientific 
Research on Priority Areas (Physics of CP violation) from the Ministry of 
Education, Science and Culture of Japan, and by the Japan-Taiwan Cooperative
Program of the Interchange Association.

\newpage

\begin{table}[tbH]
\begin{center}
{Table 1: Track helix parameter resolutions
compared with Monte Carlo predictions for
perfect alignment}
\vspace{1mm}
\begin{tabular}{|l|c|c|c|c|c|}
\hline  
  & $d\rho(\mu{\rm m})$ &$\phi_0$ (mrad)&$\kappa(10^{-3})$&
dz$(\mu{\rm m})$& tan$\lambda(10^{-3})$\\
\hline
Data&28.0&0.747&2.63&42.1&0.973\\
\hline
Perfect (MC)&19.4&0.503&1.73&34.4&0.891\\
\hline
\end{tabular}
\end{center}
\end{table} 
.
\par

\newpage

\begin{figure}[tbh]
\begin{center}
\includegraphics[scale=0.7,clip]{aiharafig8.epsi}
\caption{(a)  gain versus radiation dose and 
(b) equivalent noise count (ENC) versus radiation dose 
for AMS $0.8~\mu$m process VA1 chip.
\label{va08}}
\end{center}
\end{figure}   

\begin{figure}[tbh]
\begin{center}
\includegraphics[scale=0.9,clip]{aiharafig9.epsi}
\caption{(a)  gain versus radiation dose and 
(b) equivalent noise count (ENC) versus radiation dose 
for AMS $0.35~\mu$m process VA1 chip.
\label{va035}}
\end{center}
\end{figure}   


\begin{thebibliography}{999}

\bibitem{Belle}[1] Belle Collaboration, Technical Design Report, KEK Report 95-1, 1995.

\bibitem{KEKB}[2] KEKB B Factory Design Report, KEK Report 95-7, 1995.

\bibitem{va1}[3] E.~Nygard {\it et al.}, Nucl. Instrm. and Meth.
{\bf A301}(1991)506. 
 O.Toker {\it et al.} Nucl. Instrm. and Meth. {\bf A340}(1994) 572.

\bibitem{fadc}[4] M.~Tanaka {\it et al.}, Nucl. Instrm.and Meth. {\bf A432} (1999)422.

\bibitem{ta} [5] H.~Aihara, ``Physics Results from Belle,''
Proceedings of
the 3rd International Conference on B Physics and CP Violation,
December 3-7, 1999, Taipei, Taiwan.

\bibitem{iijima} [6] T.~Iijima, ``Kaon Identification in Belle,'' {\it ibid.}

\bibitem{yo} [7] M.~Yokoyama, 
``Test results of VA1 fabricated by sub-micron technology,'' 
This proceedings.

\end{thebibliography}
\end{document}